\documentclass[10pt,prl,aps,twocolumn,showpacs]{revtex4-1}
\usepackage{graphicx}
\usepackage{amsmath}
\usepackage{amssymb}
\usepackage{bm}
\usepackage{dcolumn}
\usepackage{subfigure}
\usepackage{psfrag}
\usepackage{float}
\usepackage{subeqnarray}
\newcommand{\comment}{\normalfont}

\begin{document}

\title{Unconventional vortex states in nanoscale superconductors due to shape-induced resonances in
the inhomogeneous Cooper-pair condensate}

\author{L.-F. Zhang}
\author{L. Covaci}
\author{M. V. Milo\v{s}evi\'{c}}
\author{G .R. Berdiyorov}
\author{F. M. Peeters}\email{francois.peeters@ua.ac.be}
\affiliation{Departement Fysica, Universiteit Antwerpen,
Groenenborgerlaan 171, B-2020 Antwerpen, Belgium}

\begin{abstract}
Vortex matter in mesoscopic superconductors is known to be strongly
affected by the geometry of the sample. Here we show that in
nanoscale superconductors with coherence length comparable to the
Fermi wavelength the shape resonances of the order parameter results in an
additional contribution to the quantum topological confinement - leading
to unconventional vortex configurations. Our Bogoliubov-de Gennes
calculations in a square geometry reveal a plethora of asymmetric,
giant multi-vortex, and vortex-antivortex structures, stable over a
wide range of parameters and which are very different from those predicted by the
Ginzburg-Landau theory. These unconventional states are
relevant for high-T$_c$ nanograins, confined Bose-Einstein
condensates, and graphene flakes with proximity-induced
superconductivity.
\end{abstract}

\maketitle
In the last decades, the effect of the boundary on mesoscopic
superconductors with dimensions comparable to the penetration depth
$\lambda$ or the coherence length $\xi$ has been intensively studied
\cite{meso1, meso2, meso3, meso4, meso5, meso6, meso7, meso8}. In applied magnetic field, it was
found that the vortex states strongly depend on the size and
geometry of the sample and are generally different from the
Abrikosov triangular lattice observed in bulk type-II conventional
superconductors (where only the vortex-vortex interaction plays a
role). For example, a giant vortex induced by strong boundary
confinement was predicted \cite{meso2} as the ground state in disks
which was subsequently observed experimentally \cite{gv1,gv2}. In
square samples, a peculiar state with an anti-vortex at the center
surrounded by four vortices was predicted theoretically \cite{anti,
anti2}, but never observed experimentally up to now.

All of the above theoretical works are based on the Ginzburg-Laudau
(GL) theory. When the
superconductor is downscaled to nano-meter sizes, quantum
confinement \cite{Anderson} leads to unique phenomena, especially in
samples with dimensions of the order of the Fermi wave length
$\lambda_F$. The GL theory is no longer applicable in this regime and the microscopic
Bogoliubov-de Gennes formalism is required. The discretization of the energy levels around
the Fermi level $E_F$ was shown to induce quantum-size effect
\cite{qse1, qse2}, quantum-size cascades \cite{cascades} and the
shell effect \cite{shell}. As one of the important results, Ref.
\cite{qse2} reported the wave-like inhomogeneous spatial
distribution of the order parameter, further enhanced at the
boundary due to quantum confinement. The latter is important
since it is well known that vortices tend to migrate and be pinned
in areas where superconductivity is suppressed \cite{inrmp}, i.e. it
is energetically favorable for a vortex to suppress the
superconducting order parameter in the region where it is already
weak. In reality the behavior is much more complex and in some instances the
vortex can be pinned where the gap is large \cite{pin}. The
appearance of oscillations in the order parameter profile due to
quantum confinement is thus expected to influence the vortex states.
For conventional superconductors, $k_F\xi_0 \approx
10^3$ ($k_F$ is the Fermi wave vector and $\xi_0$ is the BCS
coherence length), systems of size comparable to $\lambda_F$ will
not be large enough to host a vortex (being much smaller than the
coherence length). However, materials with small coherence lengths,
e.g. {\it high-T$_c$ cuprate superconductors}, will have $k_F\xi_0
\approx 1-4$ and therefore in such systems it is possible to obtain
vortex states in the quantum confinement regime. Another such
system is a {\it graphene flake} deposited on top of a
superconductor. Because of the proximity effect, superconducting
correlations will diffuse in graphene\cite{gr1, gr2, gr3, gr4}. Such a system is in the clean
limit since the scattering length in graphene is large. More
importantly, in graphene, near the Dirac point, the Fermi wavelength
is very large and can be easily manipulated by doping. In other
words, $k_f \xi_0$ can be tuned, which will allow for different
vortex patterns to be realized in the graphene flake in the quantum
confinement regime, but for more accessible sample sizes (above
100nm).

In order to experimentally detect vortex states in nano-sized
superconductors, one can extract information about the local density
of states (LDOS) from measurements of the differential conductance
with scanning tunneling microscopy (STM) \cite{gv2, exdos1, exdos2}.
An extensive analysis of the LDOS profile of the vortex states has
been performed in the past \cite{dos1, dos2, dos3, dos4, dos6}. It
is generally known that the bound states in the vortex core lead to
peaks in the LDOS at {\comment energies below the superconducting gap}, though the exact
formation of peaks in the spectrum of a multiple flux line (giant vortex) will depend on
the vorticity \cite{benxu}. {\comment Also, when $k_F\xi_0$ is small, the spectrum becomes
particle-hole asymmetric and the lowest vortex bound state has a finite energy \cite{dos1}}. In the
quantum confinement regime, there exist strong vortex-vortex and vortex-boundary
interactions and the quasiparticle spectrum becomes much more
complicated. In this case the {\comment lowest bound state} peak position does not
generally coincide with the vortex core \cite{dos5}. Furthermore,
in case of strong interactions, vortex and surface bound states may
combine to form a complex state where LDOS contributions of individual
constituents are not clearly visible.

In this Letter we report novel vortex states that appear from the interplay between quantum
confinement, inhomogeneous superconductivity, an external magnetic field,
and the sample geometry, in a nano-sized superconducting square. We
performed calculations for a sample in the quantum limit by solving
Bogoliubov-de Gennes (BdG) equations self-consistently. In what
follows, we keep constant the size of the sample and the bulk
coherence length $\xi_0=\hbar v_F /\pi\Delta_0$ (where $v_F$
is the Fermi velocity and $\Delta_0$ is the order parameter at zero
temperature), while we change the parameter $k_F\xi_0$ and thereby
tune the influence of the confinement on the vortex structure. We
start from the well-known BdG equations:
\begin{eqnarray}
\label{BdG 1}  \left[K_0-E_F\right] u_n(\overrightarrow{r})+\Delta(\overrightarrow{r})v_n(\overrightarrow{r})&=&E_nu_n(\overrightarrow{r}), \\
\label{BdG 2}  \Delta(\overrightarrow{r})^\ast u_n(\overrightarrow{r})-\left[K_0^\ast-E_F\right]
v_n(\overrightarrow{r})&=&E_nv_n(\overrightarrow{r}),
\end{eqnarray}
where
$K_0=(i\hbar\nabla+e\overrightarrow{A}/c)^2/2m$
is the kinetic energy and $E_F$ is the Fermi energy, $u_n$($v_n$)
are electron(hole)-like quasi-particle eigen-wave functions,
$E_n$ are the quasi-particle eigen-energies, and
$\overrightarrow{A}$ is the vector potential (we use the gauge
$\nabla\cdot\overrightarrow{A}=0$).

The pair potential is determined self-consistently from the
eigen-wave functions and eigen-energies:
\begin{equation}\label{OP}
\Delta(\overrightarrow{r})=g\sum\limits_{E_n<E_c}u_n(\overrightarrow{r}) v^\ast_n(\overrightarrow{r})[1-2f_n],
\end{equation}
where $g$ is the coupling constant, $E_c$ is the cutoff energy, and
$f_n=[1+\exp(E_n/k_BT)]^{-1}$ is the Fermi distribution function,
where $T$ is the temperature. We consider the two-dimensional
problem and assume a circular Fermi surface.
The confinement imposes Dirichlet boundary
conditions (i.e. $u_n(\overrightarrow{r})=0,\; v_n(\overrightarrow{r})=0,\; r \in \partial S$)
such that the order parameter vanishes at the surface. In an extreme
type-II superconductor (and/or very thin sample), it is reasonable
to neglect the contribution of the supercurrent to the total
magnetic field. For such a case, {\comment we discretize Eqs. (\ref{BdG
1}-\ref{OP}) and by using the finite difference method we solve them self-consistently.}

The free energy \cite{FreeEg1, FreeEg2} of the system is then
calculated as:
\begin{eqnarray}\label{FE}
F&=&\sum\limits_{n}2E_nf_n  + k_BT[f_n\ln f_n+(1-f_n)\ln(1-f_n)] \nonumber \\
&+&\int d\overrightarrow{r} \left[-2\sum\limits_{n}E_n|v_n|^2 + 2\Delta(\overrightarrow{r})\sum\limits_{n}u^{*}_n v_n[1-2f_n] \nonumber \right. \\
&-&\left. g\sum\limits_{n}u^*_nv_n(1-2f_n) \sum\limits_{n'}u_{n'}v^\ast_{n^\prime}(1-2f_{n^\prime}) \right],
\end{eqnarray}
where the spatial dependence of $u_n$ and $v_n$ is implicit. The
local density of states $N(r,E)$ is calculated from
\begin{equation}\label{LDOS}
N(r,E)=-\sum\limits_{n} [f'(E_n-E)|u_n|^2+f'(E_n+E)|v_n|^2].
\end{equation}

In this paper, we consider as an example a thin superconducting square with size
$5\xi_0 \times 5\xi_0$. The microscopic parameters are set to keep
$\Delta_0/E_c=0.2$. The calculations are done for
different parameters $k_F\xi_0$. Since we consider the zero
temperature case, the system is always in the quantum limit (where
$T<1/k_F\xi_0$).
%
\begin{figure}[ttt]
\includegraphics[width=\columnwidth]{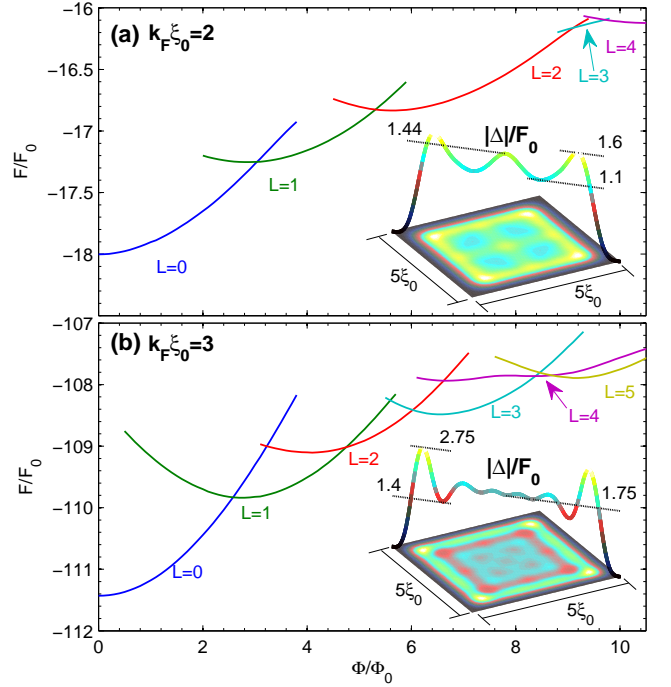}
\caption{ (Color online) Free energy as a function of the magnetic flux through the
square sample, for (a) $k_F\xi_0=2$, and (b) $k_F\xi_0=3$.  Here, $F_0=\hbar^2/2m\xi_0^2$.  The
insets show the contour plots of the order parameter with the diagonal profiles in the absence
of applied magnetic field.} \label{fig.1}
\end{figure}
%
Fig. \ref{fig.1} shows our numerical results for the free energy of
the found stable vortex configurations (the states with up to five
vortices are shown) for two values of the key parameter, $k_F \xi_0$.
The insets show
the inhomogeneous profile of the superconducting order parameter
{\it in the absence of an applied magnetic field} which is expected to
strongly influence the vortex structure. When comparing with
conventional free energy curves obtained from the GL theory
\cite{bjb}, many differences can be observed. First, the
penetration field for the first vortex is suppressed because the
order parameter is not homogeneous, allowing the vortex to penetrate
easier at locations where the order parameter is weakened. Second,
the stability range in flux for different vortex states (with
vorticity $L$) is not monotonically decreasing towards 1$\Phi_0$ as
$L$ increases. Moreover, those stability ranges
strongly vary when $k_F \xi_0$ is changed! For example, for $k_F
\xi_0=2$ the vortex structures with even vorticity are stable over a
broader magnetic field range while for $k_F \xi_0=3$ surprisingly the structures
with odd vorticity are the favored ones! The main reason behind this
phenomenon is that different confinement-induced oscillations in the
order parameter for different $k_F \xi_0$ value will stabilize
different symmetries of the vortex pattern. To illustrate this
effect better, we plot in Fig.~\ref{fig.t} the applied magnetic flux
at which ground-state transitions between states with consecutive
vorticities occur, as a function of $k_F\xi_0$. Notice the varying ranges of stability of different vortex
states, which are very sensitive to $k_F\xi_0$. Of course, for large
$k_F\xi_0$ the behavior of the system converges to a more
conventional picture (with each new vortex entering the system with
roughly one flux-quantum added).
%
\begin{figure}[ttt]
\includegraphics[width=\columnwidth]{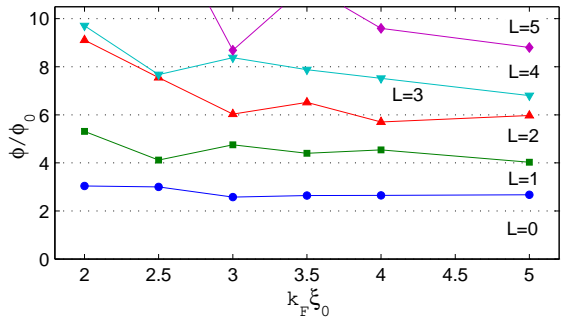}
\caption{(Color online) Transition fluxes in units of $\Phi_0$ between ground states with
consecutive vorticities for different $k_F\xi_0$.} \label{fig.t}
\end{figure}
%

\begin{figure}[ttt]
\includegraphics[width=\columnwidth]{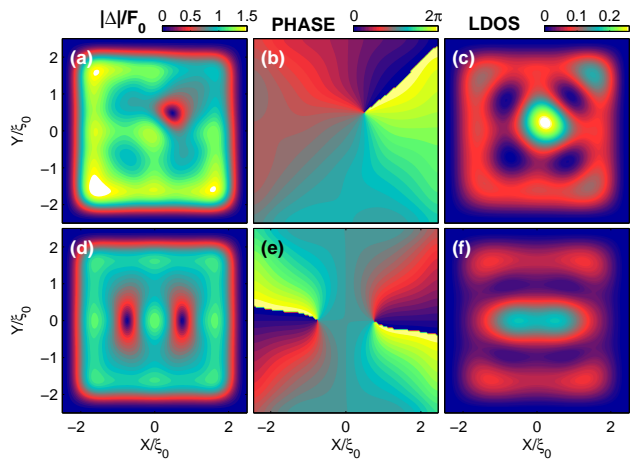}
\caption{(Color online) Contour plots of the absolute value of the
order parameter (left), the phase of the order parameter (center)
and {\comment the LDOS at $E/F_0=0.57$ in (c) and $E/F_0=0.48$ in (f)} (right) for $k_F\xi_0=2$.
Panels (a), (b) and
(c) are for $\Phi/\Phi_0=4$ and $L=1$ state. Panels (d), (e) and (f)
correspond to $\Phi/\Phi_0=6$ and $L=2$ state. } \label{fig.2}
\end{figure}

To underpin the reasons for this varying stability of vortex states
in what is otherwise a rather simple, square system, we show in
Fig.~\ref{fig.2} some of the typical states for the case of
$k_F\xi_0=2$ (the order parameter, its phase, and corresponding
LDOS). As elaborated above, the quantum confinement of electrons
here strongly affects the spatial distribution of the order
parameter [see inset in Fig.~\ref{fig.1}(a)], having three
oscillations across the square and four distinct minima that enhance
the fourfold symmetry.  This automatically leads to the
improved stability of states with even vorticity, similar to the
case of a square with 4 antidots \cite{golibPRB}. We also observe
that, before ceasing at the boundary, the order parameter is
enhanced near the surface, with the highest value found near the
corners. Due to the effect of the boundary and the shape resonances, the order parameter is also enhanced at the center of the square . To
reiterate a fairly obvious point, vortices are repelled by the peak
positions of the superconducting pair amplitude, and the four low
amplitude locations (with value only $2/3$ of the peaks) will pin
vortices rather strongly. Figs.~\ref{fig.2}(a-c), show the $L=1$
ground state for applied flux $\Phi/\Phi_0=4$. Surprisingly,
we find that the only vortex in this state is actually sitting in
one of the minima of the order parameter and the {\it fourfold
symmetry is broken}. We emphasize that this state is {\it not possible} within the GL formalism where the
single vortex will always sit in the center of the square.

{\comment From an experimental point of view, in the absence of any reference energy, the
zero-bias LDOS is most relevant. Here instead we will show the LDOS for the lowest energy vortex
bound state, which for an isolated vortex could be found from a simple empirical formula
$E_{low}/\Delta=(2k_F\xi_0)^{-1}\ln(3.33 k_F\xi_0)$ \cite{dos1}.
Therefore for the $L=1$ state we plot the LDOS at $E/F_0=0.57$}
[Fig.~\ref{fig.2}(c)]. {\comment Note that this lowest vortex bound state is not localized in the
vortex core but is shifted towards the center of the square.} We attribute this to the interactions
of the quasiparticles not only with the four deepest minima of the
inhomogeneous order parameter but also with the edges and the corners -
where the order parameter is also suppressed [see inset of Fig.~\ref{fig.1}(a)]. The effect of this
interaction can also be inferred from the finite LDOS at the corner of the sample, next to the
vortex.
%
%

When increasing further the magnetic field, an additional vortex enters
the system (forming $L=2$ state) and another unexpected spatial
distribution is stabilized. We illustrate this in
Figs.~\ref{fig.2}(d-f) for applied flux $\Phi/\Phi_0=6$. The
confinement seems to act strongly and vortices are compressed
closer to each other. However, the enhancement of the order
parameter in the center of the sample due to quantum resonance
prevents the two vortices from merging. As a consequence, vortices
are {\it squeezed into elliptical shapes}, as a pair parallel to one
of the sample edges. This vortex configuration is as different as
one can be from the known GL results, where the two vortices are {\it
always} found sitting on the diagonal, or merged into a giant
vortex, and have always an almost circular core. The LDOS plot [Fig.~\ref{fig.2}(f)] again reveals strong competing
interactions, different from those acting on vortices. For example,
we see evidence of the interaction of bound states inside the vortex
cores, since the maximum in the LDOS is reached between the vortices
and not at the center of each vortex. Also, the vortex-surface
interaction of the bound states is enhanced, leading to LDOS being
clearly appreciable near the surface.

To check further the influence of the length-scales in our sample, we also calculated the vortex states for $k_F\xi_0=3$.
In this case, the inset of Fig.~\ref{fig.1}(b)
shows six local maxima along the diagonal, twice as many as in the
$k_F\xi_0=2$ case. There are also strong oscillations near the
corners but relatively flatter away from the boundary. Again, as
expected, the four-fold symmetry is maintained, but now there are no
strong minima in the fourfold arrangement deep inside the sample. For
that reason, for $\Phi/\Phi_0=3$, the $L=1$ ground state is conventional and
contains one vortex at the center of the square. For $\Phi/\Phi_0=5.8$, the $L=2$ state is the ground state and although it still
shows the vortex pair parallel to the side of the square, the shape of the vortices and the location of the
{\comment lowest bound state} LDOS peaks bring it closer to the conventional picture
\cite{suppl2}.
\begin{figure}[ttt]
\includegraphics[width=\columnwidth]{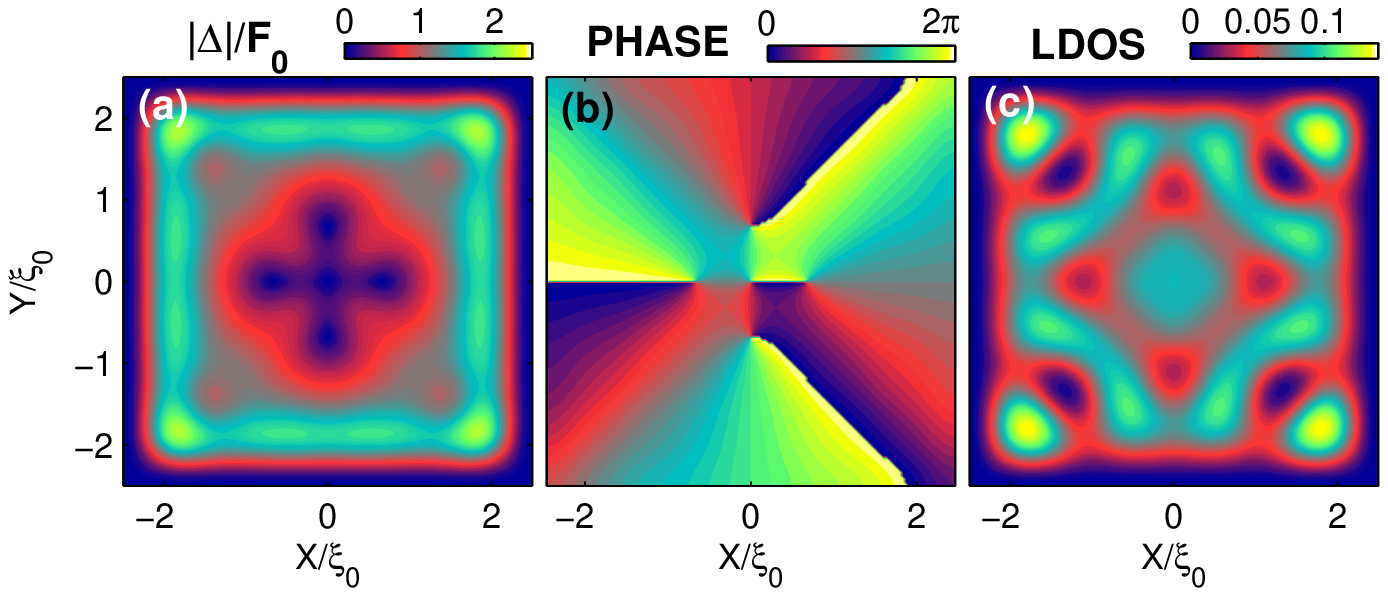}
\includegraphics[width=\columnwidth]{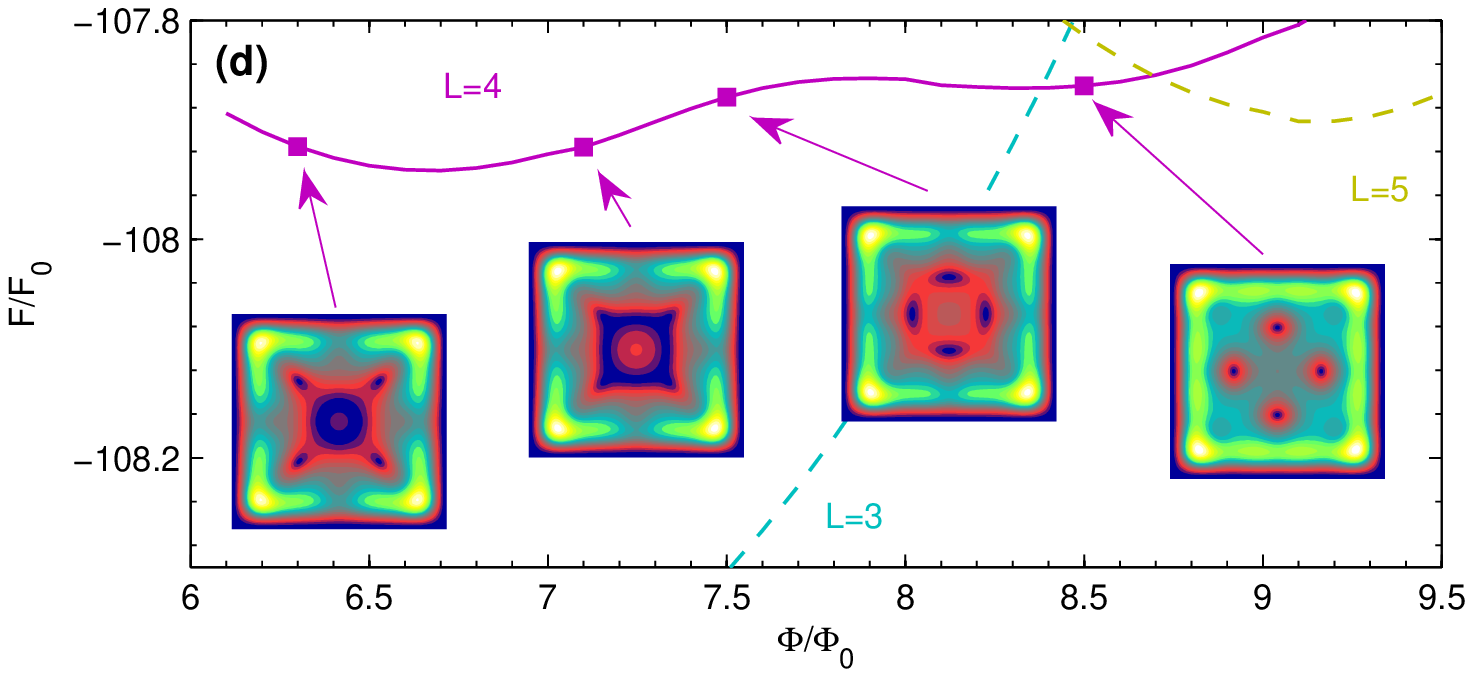}
\caption{(Color online) Contour plots of the absolute value of the
order paramenter (a), the phase of the order parameter (b)
and the LDOS at $E/F_0=0.57$ in (c) for $k_F\xi_0=3$ and
$\Phi/\Phi_0=7$ ($L=3$). Panel (d) shows the vortex configurations for the $L=4$ state and their corresponding energies.} \label{fig.3}
\end{figure}

However, when the flux through the square is increased to
$\Phi/\Phi_0=7$, as shown in Figs.~\ref{fig.3}(a-c), the ground
state has vorticity $L=3$ and is {\it not conventional}. We in fact
find the vortex-antivortex (v-av) molecule, similar to the
symmetry-induced ones predicted by the GL theory \cite{anti,
anti2} (4 vortices with an anti-vortex in between, so that the total
vorticity is $L=4-1$). There the size of the v-av molecule was found
to be very small, possibly larger if artificial pinning centers are
introduced \cite{anti2,VaV2}. In the present case, the vortex-antivortex
molecule is {\it stable over a wide range of fields} because of the symmetry of the oscillatory pattern of
the order parameter due to the quantum confinement. {\comment The LDOS at $E/F_0=0.57$}, shown in
Fig.~\ref{fig.3}(c), reveals again that
since the vortices are located closer together, the bound states are
not localized in the cores but are {\comment extended over the whole square}.

Another found difference from earlier studies is the behavior of the $L=4$ state for $k_F\xi_0=3$. As shown in the free energy curve in Fig.~\ref{fig.3}(d),
we revealed a {\it continuous phase transition} between the configuration with the vortices located on the diagonal (at lower fields)
and the configuration with the four vortices sitting near the edges of the square (higher fields), a state never found within GL!
The transition between the two four-fold symmetric states is quite peculiar and yet unseen in mesoscopic superconductivity - it involves the appearance of vortex-antivortex pairs near the center of the square \cite{suppl}. As the field is increased, the diagonal vortices annihilate with the central anti-vortices, and central vortices move to the side location. Moreover, the $L=4$ configuration with side vortices becomes the ground state for $\Phi/\Phi_0=8.5$.  Due to the inhomogeneity of the order parameter, these vortices never merge into  a giant vortex, contrary to the known GL picture for samples of smaller sizes.

In conclusion, we found {\it novel} vortex states with
{\it unconventional} stability ranges and unconventional transition between them in a superconducting square in
the quantum limit, where significant departures from previous
works based on the GL theory are found. Experimentally, these states can be accessed
through STM measurements. Additionally, we
showed that competing interactions in the quantum limit for the
bound states are different from those for the vortices, so
that the conventional picture of a vortex bound to {\comment  lowest} energy states
does not hold. {\comment Instead we predict that the maxima in LDOS of the lowest energy
states} will be observed between vortices and near surfaces. These peculiar
phenomena are made possible by strong quantum confinement, which
induces spatial oscillations in the order parameter. Their specific
pattern depends on the ratio of $\xi_0$ and $\lambda_F$, which is
unfavorable for oscillations in elementary superconductors, but is
small enough in high-$T_c$ materials. However, to observe these
novel states in the latter case, one should deal with very small
samples. As an alternative, we propose the study of a graphene flake in
contact to a superconducting film, where the Fermi energy of graphene can be tuned by a gate. In the case of graphene on Pb, our
calculations show that one could tune $k_F\xi_0$ in the
broad range of $0.1 - 10$ by shifting the Fermi energy in a
400x400nm flake from 0.01 to 0.1 eV above the Dirac point. Another
system where effects of quantum confinement on vortex matter can be
probed systematically are the optically trapped cold gases
\cite{xxx}, which are nowadays extremely controllable. Further investigations will address the rich physics in the
quantum limit, and show the effects of our findings in the
3D-confined case\cite{3d1,3d2,3d3} and multi-condensate samples\cite{2band}, but where also
barriers for vortex motion across the oscillating landscape can be investigated for,
possible use as Q-bits or other vortex devices\cite{Qbit}.

This work was supported by the Flemish Science Foundation (FWO-Vlaanderen).

\pagebreak


\begin{thebibliography}{99}
\bibitem{meso1} A. K. Geim, I. V. Grigorieva, S. V. Dubonos, J. G. S. Lok, J. C. Maan, A. E. Filippov, and F. M. Peeters, Nature (London) {\bf 390}, 259 (1997).
\bibitem{meso2} V. A. Schweigert, F. M. Peeters, and P. S. Deo, Phys. Rev. Lett. {\bf 81}, 2783 (1998).
\bibitem{meso3} J. J. Palacios, Phys. Rev. Lett. {\bf 84}, 1796 (2000).
\bibitem{meso4} B. J. Baelus, F. M. Peeters, and V. A. Schweigert, Phys. Rev. B {\bf 63}, 144517 (2001).
\bibitem{meso5} M. V. Milo\v{s}evi\'{c}, S. V. Yampolskii, and F. M. Peeters, Phys. Rev. B {\bf 66}, 024515 (2002).
\bibitem{meso6} C.-Y. Liu, G. R. Berdiyorov, and M.V. Milo\v{s}evi\'{c} , Phys. Rev. B {\bf 83}, 104524 (2011).
\bibitem{meso7} L. F. Chibotaru, A. Ceulemans, V. Bruyndoncx, and V. V. Moshchalkov, Phys. Rev. Lett. {\bf 86}, 1323 (2001).
\bibitem{meso8} G. Teniers, L. F. Chibotaru, A. Ceulemans and V. V. Moshchalkov, Europhys. Lett., {\bf 63} 296 (2003).
\bibitem{gv1} A. Kanda, B. J. Baelus, F. M. Peeters, K. Kadowaki, and Y. Ootuka, Phys. Rev. Lett. {\bf 93}, 257002 (2004).
\bibitem{gv2} T. Cren, L. Serrier-Garcia, F. Debontridder, and D. Roditchev, Phys. Rev. Lett. {\bf 107}, 097202 (2011).
\bibitem{anti} L. F. Chibotaru, A. Ceulemans, V. Bruyndoncx, and V. V. Moshchalkov, Nature (London) {\bf 408}, 833 (2000).
\bibitem{anti2} R. Geurts, M. V. Milo\v{s}evi\'{c}, and F. M. Peeters, Phys. Rev. Lett. {\bf 97},
137002 (2006). {\it ibid.}, Phys. Rev. B {\bf 75}, 184511 (2007).
\bibitem{Anderson} P. W. Anderson, J. Phys. Chem. Solids {\bf 11}, 26-30 (1959).
\bibitem{qse1} A. A. Shanenko, M. D. Croitoru, and F. M. Peeters, Phys. Rev. B {\bf 75}, 014519 (2007).
\bibitem{qse2} M. D. Croitoru, A. A. Shanenko, and F. M. Peeters, Phys. Rev. B {\bf 76}, 024511 (2007).
\bibitem{cascades} A. A. Shanenko, M. D. Croitoru, and F. M. Peeters, Phys. Rev. B {\bf 78}, 024505 (2008).
\bibitem{shell} S. Bose, A. M. Garcia-Garcia, M. M. Ugeda, J. D. Urbina, C. H. Michaelis, I. Brihuega, and K. Kern, Nat. Mater. {\bf 9}, 550 (2010).
\bibitem{inrmp} G. Blatter, M. V. Feigelman, V. B. Geshkenbein, A. I. Larkin, and V. M. Vinokur, Rev. Mod. Phys. {\bf 66}, 1125 (1994).
\bibitem{pin} D. Valdez-Balderas and D. Stroud, Phys. Rev. B {\bf 76}, 144506 (2007).
\bibitem{gr1} H. B. Heersche, P. Jarillo-Herrero, J. B. Oostinga, L. M. K. Vandersypen, and A. F. Morpurgo, Nature (London) {\bf 446}, 56 (2007).
\bibitem{gr2} H. Tomori, A. Kanda, H. Goto, S. Takana, Y. Ootuka, and K. Tsukagoshi, Physica C: Superconductivity {\bf 470}, 1492 (2010).
\bibitem{gr3} A. M. Black-Schaffer and S. Doniach, Phys. Rev. B {\bf 78}, 024504 (2008).
\bibitem{gr4} L. Covaci and F.M. Peeters, Phys. Rev. B {\bf 84}, 241401 (2011).
\bibitem{exdos1} T. Nishio, T. An, A. Nomura, K. Miyachi, T. Eguchi, H. Sakata, S. Lin, N. Hayashi, N. Nakai, M. Machida, and Y. Hasegawa, Phys. Rev. Lett. {\bf 101}, 167001 (2008).
\bibitem{exdos2} T. Cren, D. Fokin, F. Debontridder, V. Dubost, and D. Roditchev, Phys. Rev. Lett. {\bf 102}, 127005 (2009).
\bibitem{dos1} N. Hayashi, T. Isoshima, M. Ichioka, and K. Machida, Phys. Rev. Lett. {\bf 80}, 2921 (1998).
\bibitem{dos2} S. M. M. Virtanen and M. M. Salomaa, Phys. Rev. B {\bf 60}, 14581 (1999).
\bibitem{dos3} F. Gygi and M. Schluter, Phys. Rev. B {\bf 43}, 7609 (1991).
\bibitem{dos4} A. S. Melnikov, D. A. Ryzhov, and M. A. Silaev, Phys. Rev. B {\bf 78}, 064513 (2008).
\bibitem{dos6} C. Berthod, Phys. Rev. B {\bf 71}, 134513 (2005).
\bibitem{benxu} Ben Xu, M. V. Milo\v{s}evi\'{c}, Shi-Hsin Lin, F. M. Peeters, and B. Janko, Phys. Rev. Lett. {\bf 107}, 057002 (2011).
\bibitem{dos5} A. S. Melnikov, D. A. Ryzhov, and M. A. Silaev, Phys. Rev. B {\bf 79}, 134521 (2009).
\bibitem{FreeEg1} I. Kosztin, S. Kos, M. Stone, and A. J. Leggett, Phys. Rev. B {\bf 58}, 9365 (1998).
\bibitem{FreeEg2} A. Spuntarelli, P. Pieri, and G. C. Strinati, Physics Reports {\bf 488}, 111 (2010).
\bibitem{bjb}  B. J. Baelus and F. M. Peeters, Phys. Rev. B {\bf 65}, 104515 (2002).
\bibitem{golibPRB} G. R. Berdiyorov, B. J. Baelus, M. V. Milo\v{s}evi\'{c}, and F. M. Peeters, Phys.
Rev. B {\bf 68}, 174521 (2003).
\bibitem{suppl2} See supplimentary material for plots of the LDOS for vorticities $L=1$ and $L=2$
in the case when $k_F\xi_=3$.
\bibitem{VaV2} R. Geurts, M. V. Milo\v{s}evi\'{c}, and F. M. Peeters, Phys. Rev. B {\bf 79}, 174508 (2009).
\bibitem{suppl} A movie containing plots of the order paramenter and the current distribution as a function of applied field  for the $L=4$ state is provided as supplementary material.
\bibitem{xxx} S. Aubin, S. Myrskog, M. H. T. Extavour, L. J. LeBlanc, D. McKay, A. Stummer, and  J. H. Thywissen, Nature Physics {\bf 2}, 384 (2006).
\bibitem{3d1} B. Xu, M. V. Milo\v{s}evi\'{c}, and F. M. Peeters, Phys. Rev. B {\bf 77}, 144509 (2008).
\bibitem{3d2} M. M. Doria, A. R. de C. Romaguera, and F. M. Peeters, Phys. Rev. B {\bf 75}, 064505 (2007).
\bibitem{3d3} M. A. Engbarth, S. J. Bending, and M. V. Milo\v{s}evi\'{c}, Phys. Rev. B {\bf 83}, 224504 (2011).
\bibitem{2band} R. Geurts, M. V. Milo\v{s}evi\'{c}, and F. M. Peeters, Phys. Rev. B {\bf 81}, 214514 (2010).
\bibitem{Qbit} M. V. Milo\v{s}evi\'{c}, G. R. Berdiyorov, and F. M. Peeters, Appl. Phys. Lett. {\bf 91}, 212501 (2007).




\end{thebibliography}
\end{document}